\documentclass[aps, prb, showpacs, twocolumn, amsmath, letterpaper]{revtex4-2}

\usepackage{graphics}
\usepackage{booktabs}

\usepackage{amssymb}
\usepackage{bm}
\setcitestyle{super}
\usepackage{verbatim}
\usepackage{graphicx}
\setkeys{Gin}{interpolate=false}
\usepackage{float}

\usepackage[english]{babel}

\usepackage{amsmath}
\usepackage{graphicx}
\usepackage{wrapfig}
\usepackage[colorlinks=true, allcolors=blue]{hyperref}
\usepackage{array}
\usepackage{booktabs}

\begin{document}

\title{Signature of Unconventional Superconductivity in the High Temperature Normal State Resistivity}

\author{Yuchen Wu}
\author{Yiwen Liu}
\author{Wanyue Lin}
\author{Zohar Nussinov}
\author{Sheng Ran}
\affiliation{Department of Physics, Washington University in St. Louis, St. Louis, MO 63130, USA}

\begin{abstract}

Unconventional superconductivity remains one of the central unsolved problems in quantum materials, and revealing its connection to the normal state is widely believed to be key to uncovering the pairing mechanism. Previous efforts have largely focused on the temperature range immediately above the superconducting transition, where specific scattering channels—such as strange-metal transport—have been identified as sharing a possible microscopic origin with superconductivity. Here, using machine learning, we demonstrate a strong correlation between normal-state resistivity and superconductivity in Fe-based superconductors. Remarkably, the predictive information reside in a wide window of 150-300 K, far above $T_c$ of this family. We further show that the signatures of superconductivity are distributed across multiple scattering channels, which requires further theoretical investigation.

\end{abstract}

\maketitle{}


Superconductivity emerges from the normal state, inheriting its interactions and collective fluctuations. In conventional superconductors, this connection is clear: the same electron–phonon interactions that dominate normal-state resistivity also provide the pairing mechanism described by BCS theory. In unconventional materials, however, the situation is far more complex: the normal state often departs from a simple Fermi liquid and exhibits anomalous behaviors such as the pseudogap and strange-metal transport \cite{keimer2015from}. 

Most prior efforts to uncover the link between the normal state and superconductivity have focused on the narrow temperature window immediately above the superconducting transition. In this regime, short-lived Cooper pairs in the precursor superconducting phase can generate an enhanced conductivity \cite{silva2001excess,pelc2018emergence,popevcivc2018percolative,yu2019universal}, whose strength is closely tied to $T_c$. Likewise, in strange metals, the coefficient of the linear-in-$T$ resistivity has been found to scale with $T_c$, suggesting that anomalous scattering and superconducting pairing may share a common microscopic origin \cite{yuan2022scaling,greene2020strange,varma2020colloquium,jiang2023interplay}. Yet these relations remain confined to limited ranges of temperature, doping, and material classes, each tied to a single mechanism. No universal normal-state signature has been identified that reliably predicts the onset of superconductivity.

Here we take a different approach, focusing on the normal-state resistivity far beyond the immediate vicinity of $T_c$. Using machine-learning models trained on resistivity data from Fe-based superconductors, we show that the normal state far above $T_c$ might already contain information about whether a compound will superconduct and, to a moderate degree, the value of its transition temperature. Our analysis further suggests that the signatures of superconductivity are distributed across multiple scattering channels. 

We focus on Fe-based superconductors for two main reasons. First, their resistivity is intrinsically complex due to the multi-band nature and the coexistence of multiple scattering mechanisms, including electron–phonon coupling, electron–electron interactions, and spin fluctuations. While strange-metal behavior has been reported in some cases, it is often intertwined with other scattering effects and therefore less distinct than in cuprates. This richness makes Fe-based superconductors an ideal testing ground, as it avoids bias toward a single interaction channel. Second, these materials have been extensively studied under chemical substitution and applied pressure, providing a large body of resistivity data that spans both superconducting and non-superconducting samples with otherwise comparable properties, enabling meaningful comparisons. The method established here can be readily generalized to larger datasets, which is expected to further improve its predictive performance and provide deeper insight into the mechanisms of unconventional superconductivity.

\begin{figure*}[ht]
    \centering
    \includegraphics[interpolate=false,width=\linewidth]{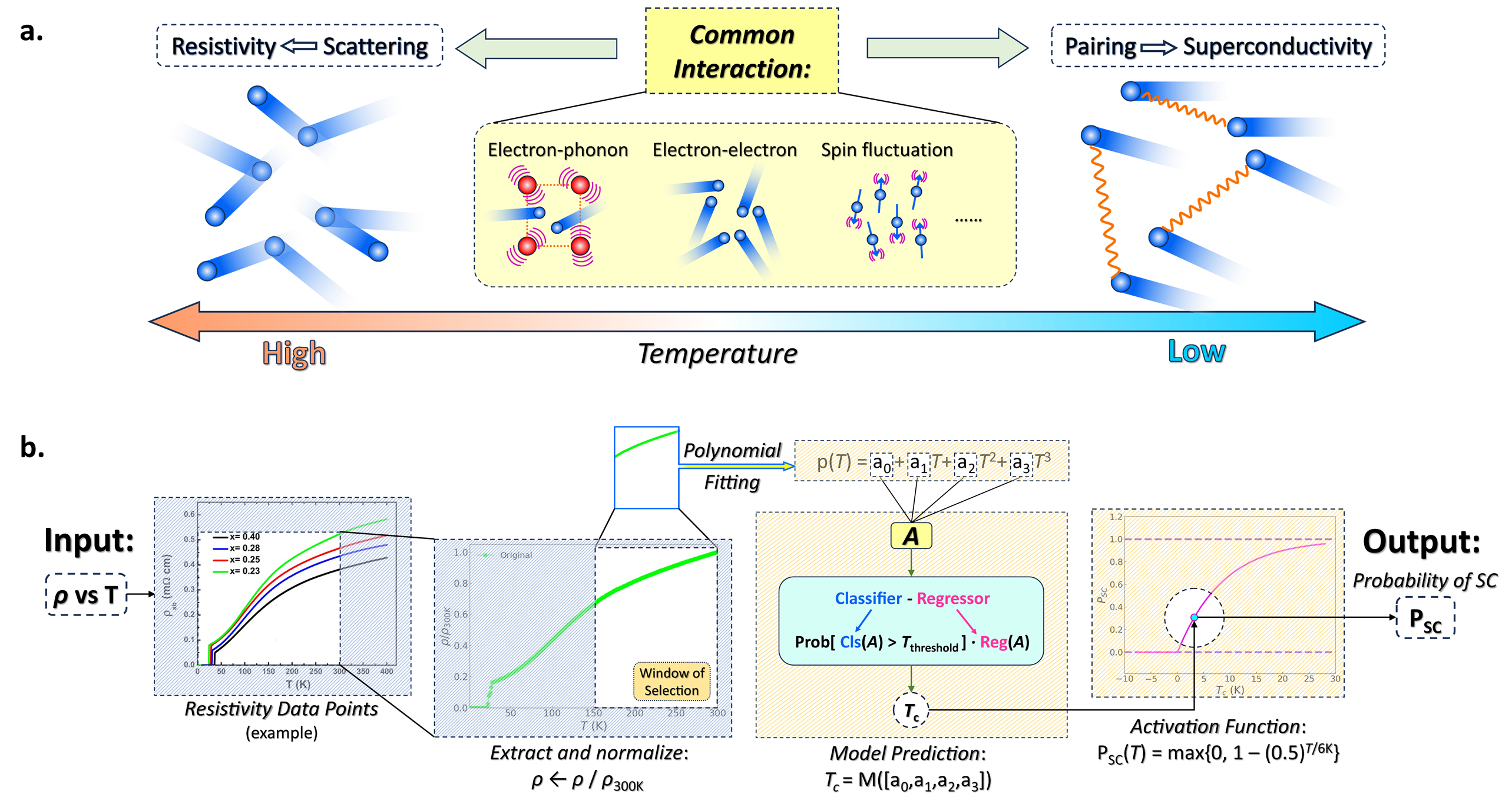}
    \caption{\textbf{Schematics of the machine learning model. a.} The interactions that govern high-temperature resistivity scattering are the same interactions that may drive superconducting pairing at low temperatures. \textbf{b.} Our machine-learning framework takes polynomial coefficients from high-temperature resistivity fits as input and predicts the superconducting critical temperature $T_c$. By applying an activation function, the regression models are further converted into classifiers to distinguish superconducting from non-superconducting samples.}
    \label{Model}
\end{figure*}

In the high temperature range, Fe-based compounds are in their paramagnetic states. At low temperatures, they will either remain in the paramagnetic state, enter magnetic ordered state, or superconducting state. Our goal is to predict the whether it enters the superconducting state, and its transition temperature $T_c$,  using resistivity data in the paramagnetic state in temperature range of ($T_{\text{min}}$, $T_{\text{max}}$), as illustrated in Fig.~\ref{Model}. Our raw dataset consists of temperature-dependent electrical resistivity curves of Fe-based superconductor family, extracted from published figures using online digitization tools. To ensure data quality, we included only resistivity curves measured on high-quality single crystals or thin films that span a full temperature range from 300 K down to 2 K (or below). In total, 175 datasets were compiled, including 115 superconducting and 60 non-superconducting samples~\cite{Li2025nickelate,
Ni2008Co,Thaler2011Mn,Ni2010NiCu,Ni2009RhPd,Thaler2010Ru,Kasahara2010P,Luo2008BaK,Torikachvili2008BaKPressure,imai2017feseTe,zhu2024la4ni3o10,kirshenbaum2010srfe2as2pt,ran2014thesis_ch6,ran2014thesis_ch7,xu2023thesis_ch5_sec55_pdf119,hou2023review11,kim2013thesis_ch4,kim2013thesis_ch5,kim2013thesis_ch6,sun2016FeTeSe_interstitialFe,Sterpetti2017BSCCO, Cai_2023}.  
Instead of feeding the raw resistivity curves directly into the models, we first fit each dataset with third-order polynomials and use the resulting coefficients as input features (details in SI). This approach suppresses noise and digitization artifacts, reduces dimensionality of input data, and provides features that are physically interpretable in terms of different scattering contributions. We trained two hybrid regression models consisting of a classifier and a regressor based on linear regression and random forest.

\begin{figure*}[ht]
    \centering
    \includegraphics[interpolate=false,width = \linewidth]{testingdataset.png}
    \caption{\textbf{The performance of the two models on the testing dataset.} In each sub-figure, the bigger plot on the top is the predicted $T_c$ plotted against the true $T_c$ of testing data; the labels are the mean and median of predicted $T_c$ (over 100 random splits of training and testing datasets) for each unique value of true $T_c$. The smaller plot on the bottom left is the distribution of the model's accuracy score over 100 random splits of training and testing datasets. The plot on the bottom-right is a snapshot of the confusion matrix of the model. These plots are made for each model: \textbf{a.} Linear Regression + Logistic Regression; \textbf{b.} Random Forest Regression + Random Forest Classification.}
    \label{Fe}
\end{figure*}

\begin{table*}
\centering
\begin{tabular}{lccccccc}
\hline
\textbf{Model} & \textbf{R$^2$ score} & \textbf{RMSE} & \textbf{AUROC} & \textbf{Accuracy} & \textbf{Std accuracy} & \textbf{Sensitivity} & \textbf{Specificity} \\
\hline
LR-based & 0.22 & 10.6 & 0.82 & 0.73 & 0.07 & 0.83 & 0.55 \\

RF-based    & 0.44 & 9.1 & 0.86 & 0.80 & 0.06 & 0.87 & 0.68\\
\hline
\end{tabular}
\caption{\textbf{Comparison of model performance metrics.} The coefficient of determination $R^2$ measures the fraction of variance in the true values explained by the predictions (with 0 indicating no correlation and 1 indicating perfect correlation). The root mean squared error (RMSE) quantifies the magnitude of the prediction error. The area under the receiver operating characteristic curve (AUROC) assesses the ability to discriminate between superconducting and non-superconducting samples (with 0.5 indicating no predictive power and 1 indicating perfect discrimination). The accuracy measures the fraction of correct predictions. The sensitivity measures the fraction of true positives correctly identified. The specificity measures the fraction of true negatives correctly identified.}
\label{metrics}
\end{table*}

Our main results, obtained with $T_{\text{min}} = 150$ K and $T_{\text{max}}$ = 300~K, are summarized in Fig.~\ref{Fe}. The predicted transition temperatures (vertical axis) are plotted against the true values (horizontal axis) for 100 runs with random train–test splits, with the diagonal line $y=x$ marking perfect agreement. For non-superconducting samples, true values of $T_c$ were set to zero. For each model, we display the mean, median, and the interquartile range containing 75$\%$ of the data. To obtain reliable statistics, we evaluated the classification accuracy—defined as the fraction of correct predictions in the test data—over 1,000 random splits, and present the resulting histograms for each model along with a representative confusion matrix.

The other performance metrics obtained from 1,000 random train–test splits are summarized in Table~\ref{metrics} for each model: the coefficient of determination $R^2$, which measures the fraction of variance in the true values explained by the predictions (with 0 indicating no correlation and 1 indicating perfect correlation); the root mean squared error (RMSE), which quantifies the magnitude of the prediction error; the area under the receiver operating characteristic curve (AUROC), which assesses the ability to discriminate between superconducting and non-superconducting samples (with 0.5 indicating no predictive power and 1 indicating perfect discrimination); the sensitivity, which measures the fraction of true positives correctly identified; and the specificity, which measures the fraction of true negatives correctly identified.

The linear regression (LR) based model serves as a baseline, fitting a single linear relation to the data. The predicted $T_c$ values collapse into a narrow range around $\sim$15 K, largely independent of the true $T_c$, leading to a low $R^2$ of 0.22, which indicates a weak correlation between predicted and actual transition temperatures. In contrast, the same model achieves a classification accuracy of 73~$\%$ and an AUROC of 0.82, demonstrating a surprisingly strong ability to distinguish superconducting from non-superconducting samples despite its poor quantitative prediction of $T_c$.

The random-forest (RF) based model exhibits improved performance. The predicted $T_c$ values follow the experimental values slightly more closely and align better with the $y=x$ line in Fig.~\ref{Fe}, although noticeable scatter remains. This results in an $R^2$ approximately twice that of the LR model. Meanwhile, both the classification accuracy and AUROC exceed 0.8, indicating robust classification capability. In both LR and RF models, the classification performance is significantly stronger than the prediction of $T_c$, which is intrinsically more challenging. The superior performance of the RF model over linear regression suggests that the signatures of superconductivity are encoded in nonlinear combinations of features, rather than in any single linear trend.

\begin{figure*}[ht]
  \centering
\includegraphics[interpolate=false,width=\linewidth]{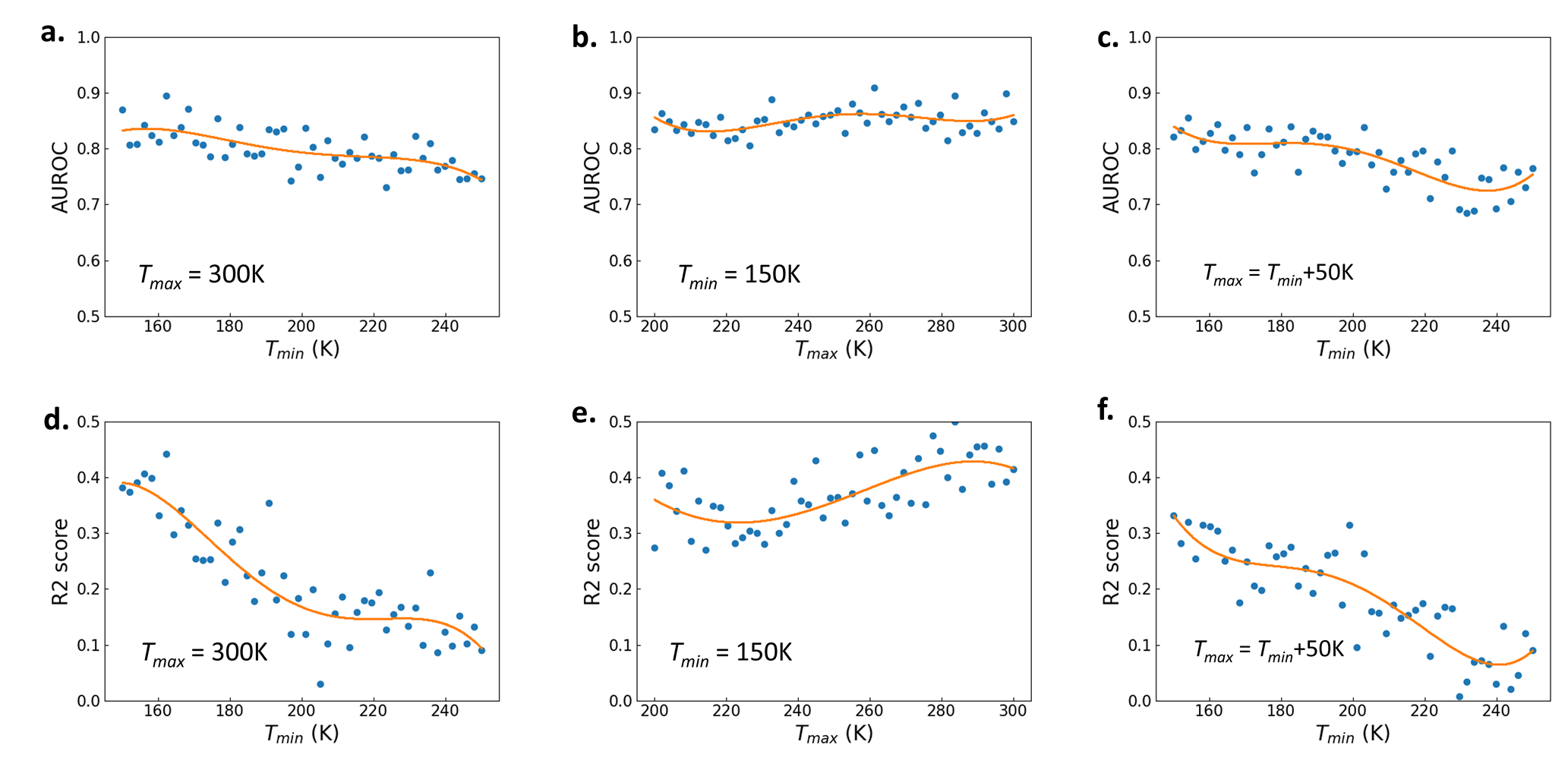}
  \caption{(a) - (c) The changes in AUROC score with varying range of temperature used by the model. (a) fix $T_{max}$ = 300K, vary $T_{min}$ (b) fix $T_{min}$ = 150K, vary $T_{max}$ (c) translation of the temperature range with $T_{max} = T_{min}+$50K. (d) - (e) The changes in R2 score with varying range of temperature used by the model. (d) fix $T_{max}$ = 300K, vary $T_{min}$ (e) fix $T_{min}$ = 150K, vary $T_{max}$ (f) translation of the temperature range with $T_{max} = T_{min}+$50K.  }
  \label{Trange}
\end{figure*}

To gain more insight into the physical origin of the predictive power of our model, we systematically varied the temperature ranges used for training and monitored both AUROC and $R^2$ of the RF model, which quantifies the ability to distinguish superconducting from non-superconducting samples and to predict $T_c$, respectively. First, we fixed the upper bound at $T_{max}$ = 300~K and gradually increased the lower bound $T_{min}$ starting from 150~K. As shown in Fig.~\ref{Trange}a, the AUROC remains approximately constant at around 0.8 and decreases only slightly as $T_{\min}$ increases. By contrast, $R^2$ drops rapidly, falling to about 0.1 once $T_{\min}$ exceeds 200~K. This indicates that the data in the 200–300~K range carries essentially no predictive information about $T_c$, even though classification information persists over a broader temperature range extending from roughly 150 to 300~K. Consistently, when we fix $T_{\min}=150$~K and reduce $T_{\max}$ from 300~K to 200~K, the $R^2$ value varies modestly but stays above 0.3 and the AUROC remains nearly unchanged at approximately 0.85.

To further confirm this trend, we tested sliding temperature windows of fixed width 50~K across the full temperature range. As the window moves toward higher temperatures, the AUROC decreases slightly but consistently remains above 0.7. In contrast, the coefficient of determination $R^2$ drops rapidly and falls below 0.1 once the lower bound of the window $T_{\min}$ exceeds 200~K. Taken together, these results demonstrate that while the most informative resistivity data for predicting the precise value of $T_c$ is concentrated at lower temperatures, the ability to classify whether a material is superconducting extends over a much broader temperature window, roughly spanning 150–300~K, with a slightly enhanced contribution from the 150–250~K range.

Our results contradict the conventional wisdom in the search for the pairing mechanism of unconventional superconductivity. Prior studies have largely focused on the temperature range immediately above $T_c$. In some classes of unconventional superconductors, the coefficient of the linear-in-$T$ resistivity—a hallmark of strange-metal transport—has been shown to scale with $T_c$ \cite{yuan2022scaling,greene2020strange,varma2020colloquium,jiang2023interplay}. However, this correlation is confined to a narrow temperature window just above the transition, typically below 100 K \cite{yuan2022scaling}, beyond which conventional Fermi-liquid behavior reemerges. Precursor effects have also been reported in cuprates: above $T_c$, short-lived Cooper pairs can form and break, producing an enhanced conductivity known as paraconductivity \cite{silva2001excess,pelc2018emergence,popevcivc2018percolative,yu2019universal}, whose magnitude is closely tied to $T_c$. Yet paraconductivity is likewise restricted to a very limited range, usually no more than 20~$\%$ of $T_c$ \cite{popevcivc2018percolative}. In both cases, the relation holds only for selected materials and centers on a single mechanism. By contrast, our dataset spans multiple Fe-based superconductor families, and our data reveal a strong correlation between superconductivity and the normal-state resistivity in the temperature range of 150–300 K. 

To further probe the physical origin of the predictive power, we trained models using selected subsets of polynomial coefficients obtained from the temperature-dependent resistivity fits, as summarized in Table II. Here, the coefficients of the $T^0$, $T^1$, $T^2$, and $T^3$ terms in the cubic polynomial fits of the normal-state resistivity are referred to as the zeroth-, first-, second-, and third-order terms, respectively. The zeroth-order term plays the least important role: when only this term is used, the AUROC drops from 0.86 to 0.67 and the $R^2$ value falls to nearly zero. In contrast, the first-order term carries the strongest individual predictive power—using it alone results in the smallest reduction in both AUROC and $R^2$. Importantly, however, the relevant information is not exclusively contained in the first-order term. For example, using only the second-order term yields an AUROC only slightly lower than that obtained from the first-order term. Moreover, when the first-order term is excluded, a combination of the zeroth-, second-, and third-order terms achieves better overall performance than using the first-order term alone. Taken together, these results indicate that the signatures of superconductivity are distributed primarily across the first-, second-, and third-order terms, with the first-order contribution being slightly more dominant but not singularly decisive.

\begin{table*}
\centering
\begin{tabular}{lccccccc}
\hline
\textbf{Coefficients} & \textbf{R$^2$ score} & \textbf{RMSE} & \textbf{AUROC} & \textbf{Accuracy} & \textbf{Std accuracy} & \textbf{Sensitivity} & \textbf{Specificity} \\
\hline
0,1,2,3    & 0.44 & 9.1 & 0.86 & 0.80 & 0.06 & 0.87 & 0.68\\
\hline
0    & 0.06 & 11.73 & 0.67 & 0.64 & 0.07 & 0.69 & 0.54\\
\hline
1   & 0.17 & 10.93 & 0.80 & 0.70 & 0.07 & 0.72 & 0.69\\
\hline
2    & 0.10 & 11.56 & 0.77 & 0.73 & 0.07 & 0.76 & 0.67\\
\hline
3    & 0.05 & 11.87 & 0.76 & 0.75 & 0.07 & 0.78 & 0.70\\
\hline
1,2,3    & 0.37 & 9.55 & 0.85 & 0.81 & 0.07 & 0.87 & 0.68\\
\hline
0,2,3    & 0.33 & 9.8 & 0.80 & 0.73 & 0.07 & 0.81 & 0.59\\
\hline
0,1,3    & 0.40 & 9.36 & 0.84 & 0.78 & 0.07 & 0.85 & 0.65\\
\hline
0,1,2    & 0.42 & 9.16 & 0.84 & 0.79 & 0.06 & 0.83 & 0.71\\
\hline
\end{tabular}
\caption{\textbf{Performance of models trained with selected polynomial coefficients.} The entries indicate which coefficients of the polynomial fit $\rho(T) = a_0 + a_1T + a_2T^2 + a_3T^3$ are included as input features for the model.}
\label{metrics2}
\end{table*}

Given that superconductivity in Fe-based compounds is widely believed to be unconventional, the linear-in-$T$ term correlated with superconductivity is unlikely to originate from conventional electron–phonon scattering, and may instead reflect strange-metal–like transport associated with strong correlations. The quadratic term, commonly attributed to electron–electron scattering within a Fermi-liquid framework, is not expected to be prominent at such elevated temperatures in simple metals. Likewise, a cubic-in-$T$ contribution lacks a clear microscopic origin in conventional three-dimensional systems. In this context, the combined importance of the second- and third-order terms suggests that they may arise from more complex processes, such as scattering mediated by magnetic fluctuations or other collective excitations. This interpretation points to an interplay of multiple unconventional scattering channels in the normal state, and calls for further theoretical work to elucidate their microscopic origin and connection to superconducting pairing.

To conclude, our findings uncover an unexpected temperature window in which the normal state retains signatures of superconductivity. This opens a new perspective on how transport encodes the onset of pairing. More broadly, our work establishes a new avenue for machine-learning–driven prediction of superconductors and other quantum materials. Prior studies in this field have typically relied on chemical composition and leveraged very large datasets \cite{goodall2020mlTc,oses2018datamining,zhang2019mlTc,wang2018supervisedSC,jiang2021nnSC,zhang2018compOnlyTc,chen2021mlNewSC,sanfilippo2022mlSC,sanna2020mlHydrides,hutcheon2020mlHydrides,zhang2020unsupervisedSC,zhang2021mlTcScreen,Broyles2024Uranium}. By contrast, our approach provides more direct and interpretable insights into the physical mechanisms, albeit at the cost of manually assembling a smaller and noisier dataset. It is conceivable that with larger and more systematic datasets, this method could become even more powerful. These results highlight an urgent need for community-wide efforts to build large experimental databases of physical observables, extending beyond structural information alone.

We acknowledge fruitful discussions with Shaffique Adam, Tao Ju, and Daoyi Zhu. The work at Washington University is supported by the National Science Foundation (NSF) Division of Materials Research Award DMR-2512965.




\bibliographystyle{unsrt}
\bibliography{main.bib}             

\clearpage

\end{document}